\documentclass{ws-procs9x6}

\usepackage{epsfig}
\begin{document}
\title{Gluon Evolution and Saturation
Proceedings\footnote{\uppercase{P}resented at the {\bf \uppercase{G}ribov \uppercase{M}emorial 
\uppercase{W}orkshop on \uppercase{Q}uantum \uppercase{C}hromodynamics and \uppercase{B}eyond}}}

\author{L.~D. McLerran}

\address{Brookhaven National Laboratory and Riken Brookhaven Center, \\
Physics Dept.\\ 
Upton, NY 11973\\ }

\maketitle

\abstracts{Almost 40 years ago,  Gribov and colleagues at the Leningrad Nuclear Physics Institute
developed the ideas that led to the Dokhsitzer-Gribov-Altarelli-Parisi  the Baltisky-Fadin-Kuraev-Lipatov equations.  These equations describe the evolution of the distributions for quarks and gluon inside a hadron to increased resolution scale of a probe or to smaller values of the fractional momentum of a hadronic  constituent.  I motivate and discuss the generalization required of these equations needed for high energy processes when the density of constituents is large.  This leads to a theory of saturation realized by the Color Glass Condensate
}

\section{Introduction}
About 40 years ago, Gribov and colleagues at the Leningrad Nuclear Physics Institute developed ideas that led to equations that describe the change in quark and gluon distribtutions in hadrons as a function of both the resolution size of an electromagnetic probe \cite{Gribov:1972ri} \cite{Gribov:1972rt} \cite{Dokshitzer:1977sg} \cite{Dokshitzer:1978hw} \cite{Lipatov:1974qm} \cite{Altarelli:1977zs} and as a function of the fractional momentum
of the parton in a hadron \cite{Kuraev:1977fs} \cite{Balitsky:1978ic}.  These developments led to a revolutionary change in our understanding of strong interaction physics and provided a foundation within QCD for ideas originally developed by Bjorken \cite{Bjorken:1969ja}.  

\begin{figure}[!htb]
\begin{center}
  \mbox{{\epsfig{figure=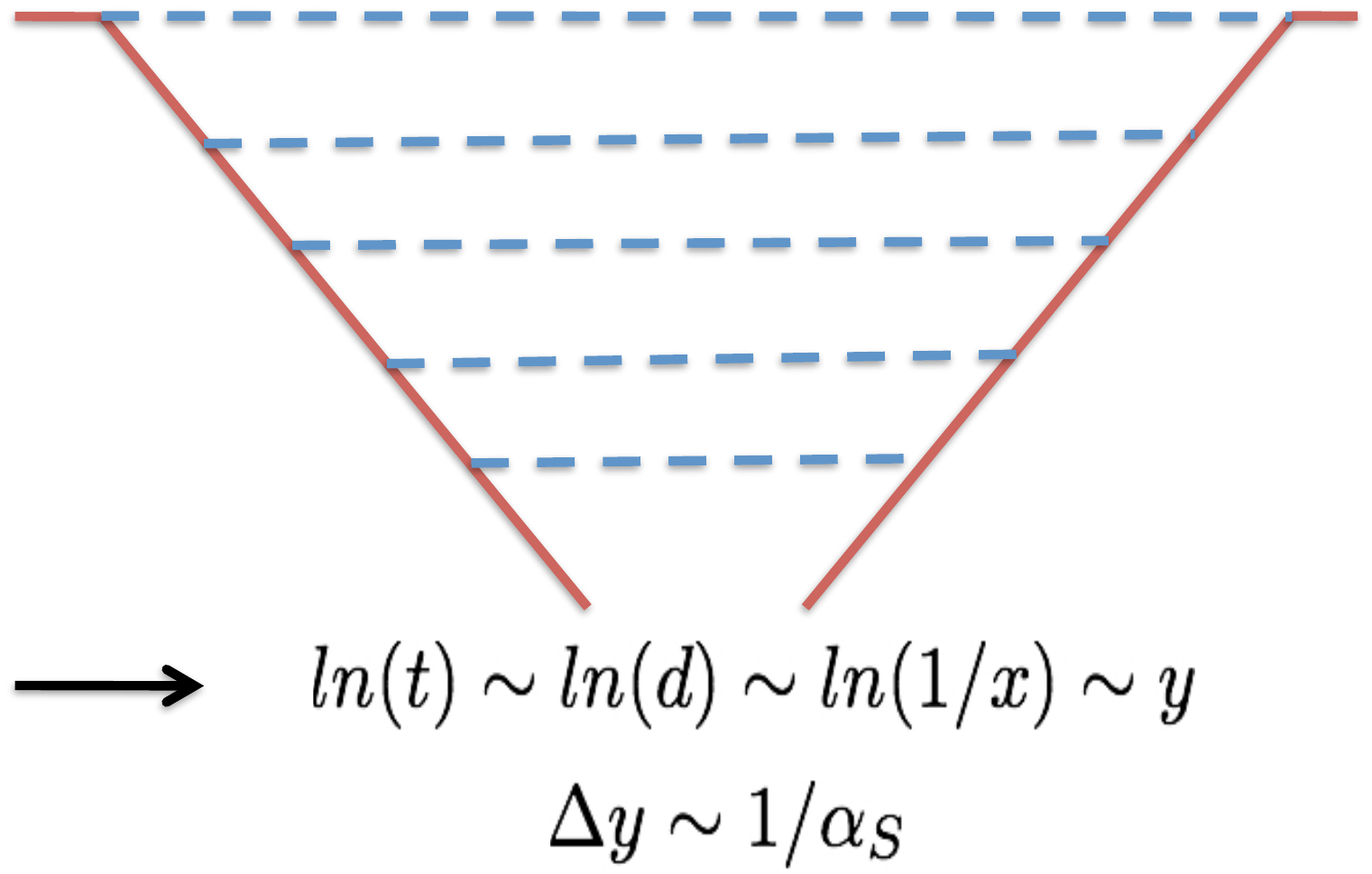, width=0.50\textwidth}}}
   \end{center}
\caption[*]{ \small  The space-time diagram that describes the evolution of a hadronic wavefunction.}
     \label{perturbation}
\end{figure}

These ideas are manifest in the space-time diagram for hadronic processes developed by Gribov and by Bjorken, as shown in Fig. \ref{perturbation}.  The red line on this diagram shows a hadronic constituent  in the initial state successively radiating and then absorbing quanta in the final state.  The position of emission in the figure corresponds to the distance either before or after the collision when the quanta is radiated.  The time is of order the distance away from the collision where the radiation takes place, and we are looking in the laboratory frame where the initial hadron has high energy.  Where the initial state and final state hadron meet, the collision takes place.  This picture formed the framework of our modern understanding of pp, pA, and AA collisions \cite{Bjorken:1976mk}.

A difficulty with the evolution equations is that they predict a rapid growth as an inverse power of the fractional momentum of the gluon distribution function.  This growth if uncontrolled can lead to violations of unitatrity in high energy collisions.  This was recognized by Leonid Gribov (Volodya Gribov's son,) Levin and Ryskin \cite{Gribov:1984tu}, and by Mueller and Qiu \cite{Mueller:1985wy} who argued that the growth should slow at fixed probe resolution scale.  Together with Raju Venugopalan, we argued that to understand such saturation of the gluon density, one needed to replace the idea of a gluon density as an incoherent distribution by the idea of a classical gluon field \cite{McLerran:1993ni} \cite{McLerran:1993ka}.  This idea leads to renormalization group equations that are generalizations of those envisioned by Gribov. These equations can include the effect of the high density of gluons, and do not violate s-channel unitarity \cite{JalilianMarian:1996xn} \cite{JalilianMarian:1997jx} \cite{JalilianMarian:1997gr} \cite{Iancu:2000hn} \cite{Iancu:2001ad} \cite{Ferreiro:2001qy}.  The region where the gluon density is large is where  the classical field nature of the gluon density is important.  The high gluon density  provides a natural infrared cutoff in the theory,
and makes possible a number of computations of processes  that were infrared divergent in the ordinary parton model.   

A momentum scale, the saturation momentum,  characterizes this matter.  The saturation momentum grows at high energies, and the strong coupling constant ultimately becomes small when
evaluated at the saturation momentum scale.  A novel space-time picture of the early stages of hadron-hadron collsions has emerged \cite{Kovner:1995ja} \cite{Kovner:1995ts} \cite{Krasnitz:1998ns} \cite{Krasnitz:1999wc} \cite{Lappi:2003bi} \cite{Lappi:2006fp}.  The high density matter inside a hadronic wave function is called the Color Glass Condensate, and the high density highly coherent matter produced in hadronic collisions is called the Glasma.

It is the purpose of this lecture to motivate and provide an intuitive motivation for the Color Glass Condensate and the Glasma.

\section{Some Qualitative Features of Parton Evolution}

We begin by introducing the parton transverse momentum and rapidity distribution $df/dyd^2p_T$,
where $y =ln(1/x)$ and $x$ is the parton momentum as a fraction of the hadron momentum in a frame where
a hadron has a very high energy.  The parton distribution function is
\begin{equation}
  xG(x,Q^2) = \int^{Q^2} ~ d^2p_T {{df} \over {dyd^2p_T}}
\end{equation}
There is need for an upper limit on this integration since the integral would weakly diverge if the limit was taken to infinity.  The value of $Q^2$ can be thought of as a resolution scale and we are counting all partons
whose size $r \sim 1/p_T$ is $r < 1/Q$.

The DGLAP evolution equation describes the change in the parton distribution as one changes the resolution scale.  Since the integral for $G(x,Q^2)$ is mildly divergent, we see that the hadron has more and more smaller constituents.  This means that the density of partons in a hadron $\rho =G/\pi R^2$ times the typical area of a constituent, $1/Q^2$ shrinks to zero $\rho/Q^2 \rightarrow 0$ as $Q^2 \rightarrow \infty$.  The evolution in $Q^2$ takes one into a short distance dilute limit.

\begin{figure}[!htb]
\begin{center}
  \mbox{{\epsfig{figure=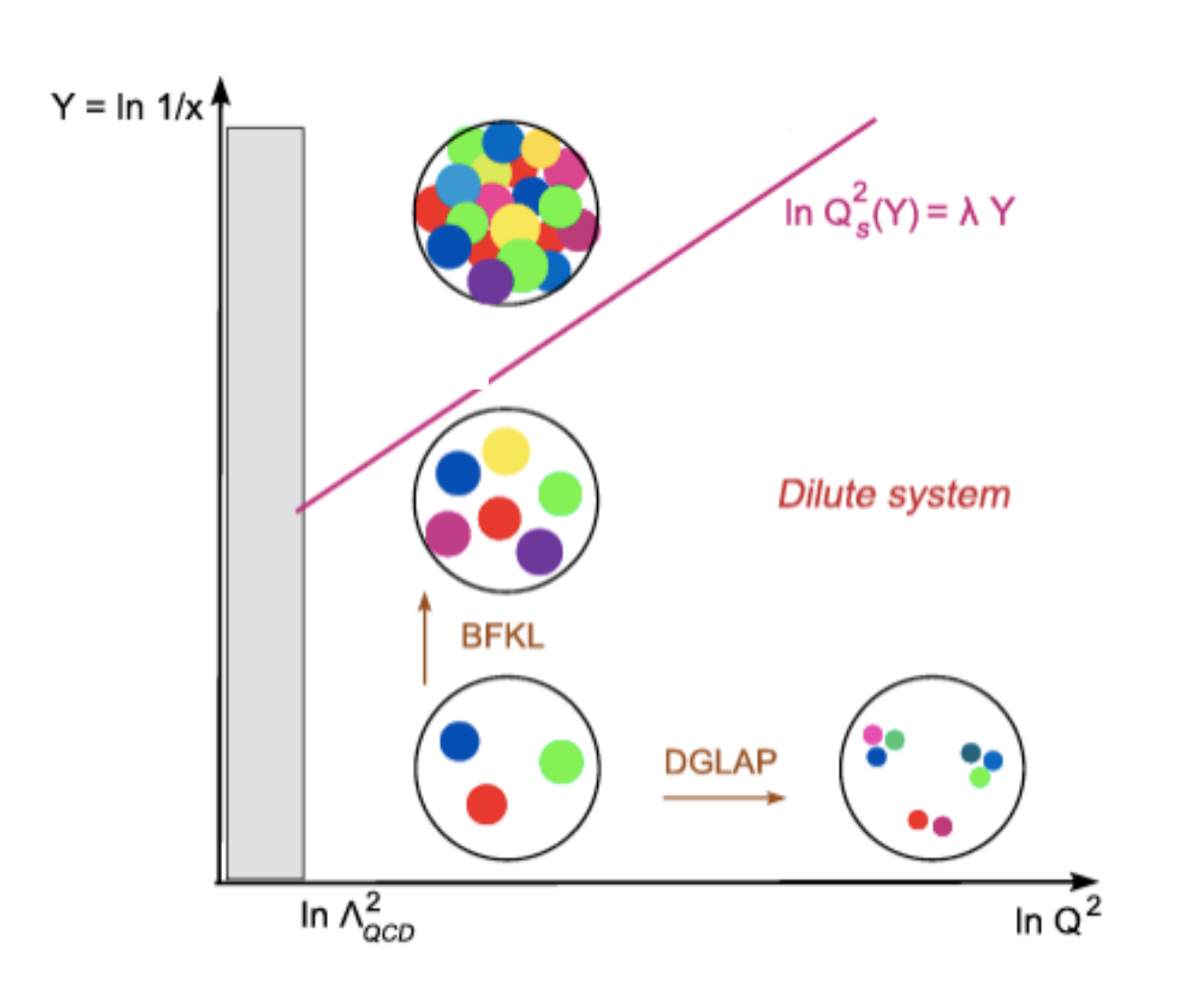, width=0.70\textwidth}}}
   \end{center}
\caption[*]{ \small  The gluon density in the $ln(x)-ln(Q^2)$ plane.}
     \label{plane}
\end{figure}

The evolution of  the gluon density in $y = ln(1/x)$ at fixed $Q^2$ is given by the BFKL equation.  The gluon density grows like $1/x^\delta$ where $\delta \sim 0.2-0.3$ at accessible energies.  Evolution at fixed $Q^2$ corresponds to evolution at fixed parton size.  Evolution in $y$ takes one to the high parton density limit.
For fixed $Q^2$ we shall soon see that the BFKL evolution equation breaks down, and the rapid growth is tempered.  A figure illustrating the behaviour of the gluon distribution function in the $ln(1/x)-ln(Q^2)$ plane is shown in Fig. \ref{plane}.  The red line in the diagram corresponds to the point when evolving
in $ln(x)$ at fixed values of $ln(Q^2)$ where the gluon density stops growing rapidly.  This line is called the saturation boundary.

\begin{figure}[!htb]
\begin{center}
  \mbox{{\epsfig{figure=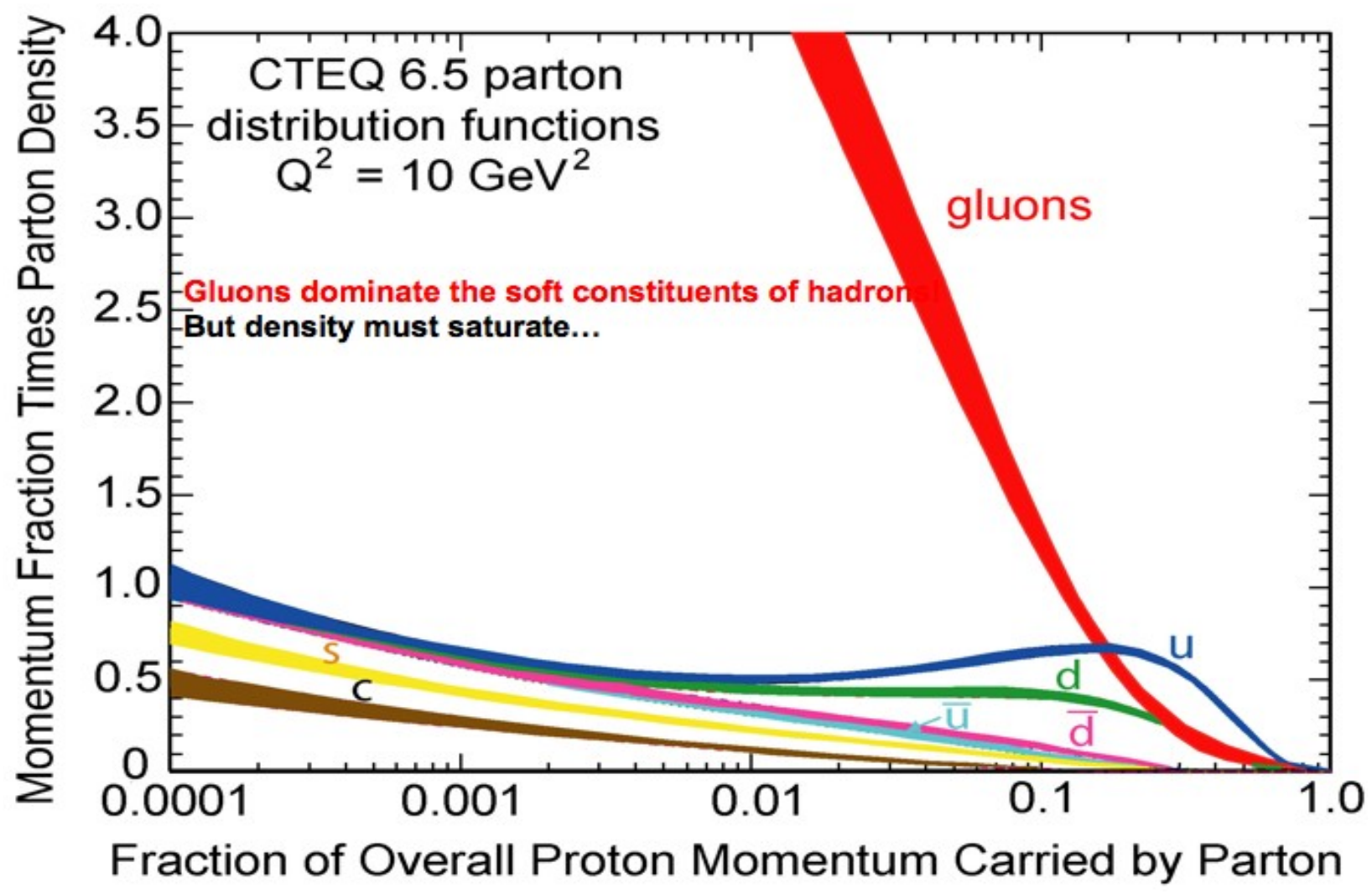, width=0.70\textwidth}}}
   \end{center}
\caption[*]{ \small  The distributions of quarks and gluons as a function of $x$.}
     \label{gluons}
\end{figure}

The growth of the gluon density is seen experimentally.  In Fig. \ref{gluons}, the gluon and quark densities are shown.  One sees that gluons dominate the density of particles inside a hadron for $x \le 10^{-2}$.  The density also is rapidly growing with decreasing $x$.  Surely when the density of gluons becomes very large,
one can use weak coupling methods to describe the gluons.  This is because of asymptotic freedom and that the typical separation between gluons  is very small.  This does not mean the system is perturbative.  The high density of gluons can act coherently and generate large interactions.  A simple example of coherence turning intrinsically weak interactions into strong forces is gravity.  This is because classical fields can add together with the same sign, and because the interaction is long range so that interactions are enhanced due to coherent forces of many nucleons.

\section{Saturation of the Gluon Density}
\begin{figure}[!htb]
\begin{center}
  \mbox{{\epsfig{figure=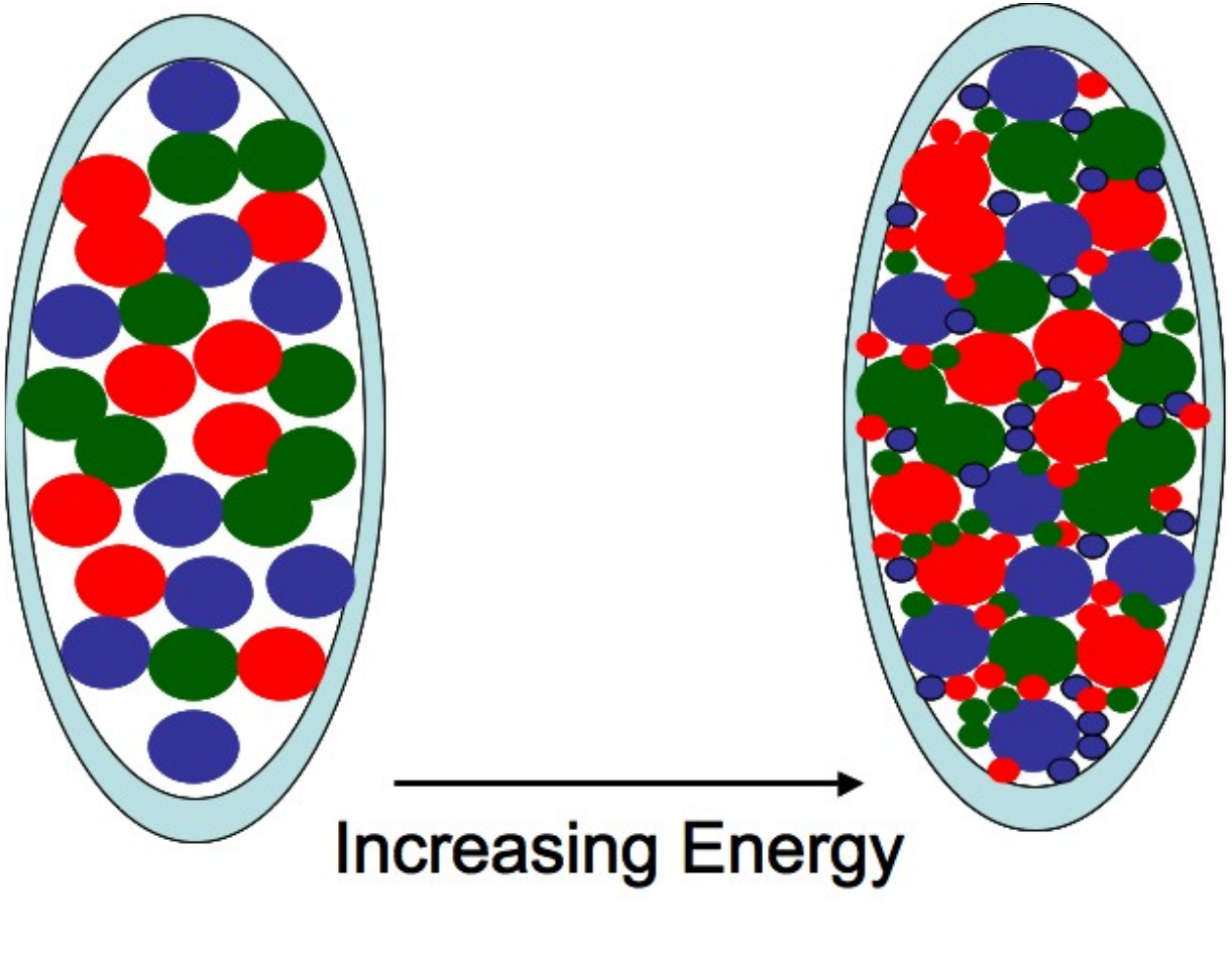, width=0.70\textwidth}}}
   \end{center}
\caption[*]{ \small  Saturation of the gluon density.}
     \label{saturation}
\end{figure}
To understand how the gluon density might saturate, imagine that the gluons are hard spheres with a size of order $r \sim 1/p_T$.  A hadron has a very slowly growing size as energy increases so we will treat the hadron size as fixed.  On the other hand, as we go to higher energy, we can probe smaller values of $x_{min} \sim \Lambda_{QCD}/E$.  If we start with a low density of gluons of size $r_0$ at some energy, the
hadronic disk begins to become closely packed with gluons of this size as the energy increases. This continues until the gluons are so closely packed that they repel, and begin to act as hard spheres.  This is the density ${{d\rho} \over {dyd^2p_Td^2r_T}} \sim 1/\alpha_S$.  At this density scale their intrinsic weak interaction strength $\sim \alpha_S$ is compensated by their high density $\sim 1/\alpha_S$.

What happens as we go to yet higher energy? We can still pack in more gluons but they have to have smaller size, $r << r_0$ so that they can fit in the cracks between the gluons of size $r_0$.  This means that the gluon density can grow forever, so long as it is associated with smaller and smaller gluons.  There is a characteristic momentum scale which at any energy scale separates the highly coherent gluons from those that are not so coherent.  This is the saturation scale $Q_{sat}$.  Our considerations argue that the saturation momentum can grow forever.  When it is $Q_{sat} >> \Lambda_{QCD}$ we can use weak coupling methods.

\begin{figure}[!htb]
\begin{center}
  \mbox{{\epsfig{figure=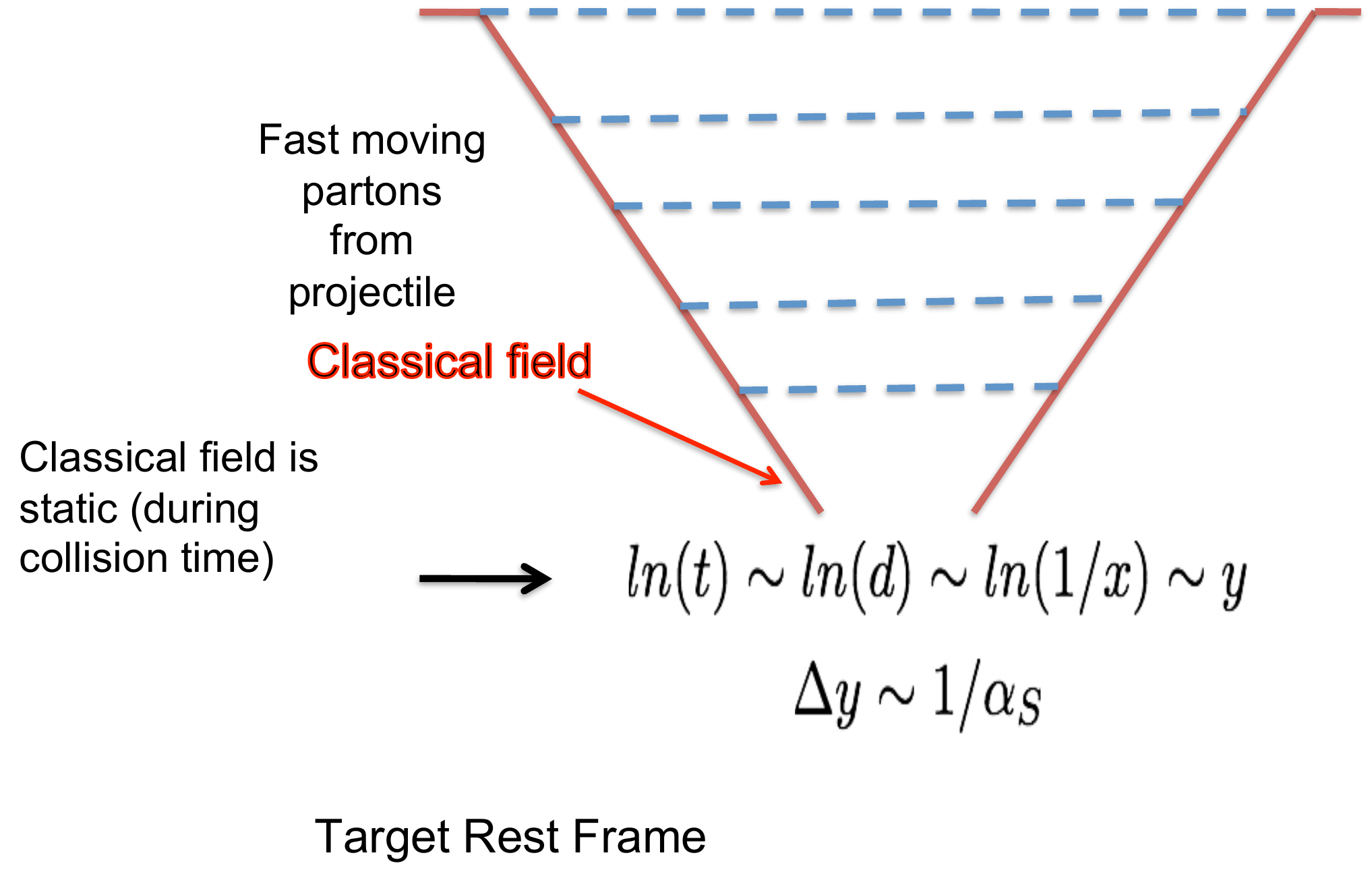, width=0.70\textwidth}}}
   \end{center}
\caption[*]{ \small  The space time diagram showing how the glassy nature of the Color Glass Condensate arises}
     \label{gribov}
\end{figure}

An essential ingredient in this description is Gribov's space-time picture shown in Fig. \ref{gribov}.  The smallest $x$ gluon emitted is the lowest on on this figure.  Because the phase space density of gluons is so large, we are justified in thinking about this gluon as a classical field.  It has however been produced by gluons at much larger $ln(x^\prime /x) \sim 1/\alpha_S$.  This gluon has its evolution Lorentz time dilated.  This means that the produced gluon field is static.  It also has a deeper consequence:  The different configurations that yield the gluon field will not quantum mechanically interfere.  They are a glass, similar to spin glasses of condensed matter physics

We therefore call this high density ensemble of gluons the Color Glass Condensate.  Color is because it is made of colored gluons.  Glass because the field describing it are static and because of its relationship
to spin glasses.  Condensate because the phase space density of gluons is very high, the gluons are highly coherent, and this high density arises spontaneously.

\section{The Renormalization Group Formalism for the CGC}

The Color Glass Condensate is described by both gluon fields and sources for them.  It has a path integral representation:
\begin{equation}
 Z = \int_\Lambda ~ [dA][d\rho]~ exp \{ iS[A,\rho] - F[\rho]\}
\end{equation}
The Yang-Mills action is in the presence of a source,
\begin{equation}
  J^\mu = \delta^{\mu+} \rho(x_T,y)
\end{equation}
The space-time rapidity is $y = ln(x_0^-/x^-)$, where $x^- = (t-z)/\sqrt{2}$.  By the uncertainty principle
$y \sim ln(1/x)$, where x is the fractional momentum of a constituent.  The scale $\Lambda$ is a separation scale that separates the gluon degrees of freedom between dynamical fields and sources.  Its presence leads to renormalization group equations that determine the distribution of gluonic sources, $F[\rho]$.  

To leading order at weak coupling, the gluon field is determined by the classical field equations.  Once this is done, one then integrates over the incoherent distribution of the sources.  Of course there are fluctuations in the gluon field, and if we were to try to compute quantities with longitudinal momentum much less that the cutoff $\Lambda$, we would generate large terms proportional to $\alpha_S ~ln(\Lambda/p^+)$.  To deal with such quantities, we need to shift the scale $\Lambda$ to be closer to $p^+$.  This is done by renormalization group methods and leads to the JIMWLK evolution equations\cite{JalilianMarian:1996xn} \cite{JalilianMarian:1997jx} \cite{JalilianMarian:1997gr} \cite{Iancu:2000hn} \cite{Iancu:2001ad} \cite{Ferreiro:2001qy}.  When computed on correlation functions, these equations are equivalent to the Balitsky-Kovchegov equations \cite{Balitsky:1995ub} \cite{Kovchegov:1999yj}.  The derivation of the JIMWLK equation requires an analytic computation to all orders in the background classical gluon field for an arbitrary light cone source.

We can understand generic features of the JIMWLK equations.  They are for the source functional
\begin{equation}
  Z_0 = e^{-F[\rho ]}
\end{equation}
They are of the form of a a Euclidean Hamiltonian evolution equation
\begin{equation}
  {d \over {dy}} Z_0 = -H[d/d\rho , \rho]~ Z_0
\end{equation}
For strong and intermediate strength fields, corresponding to the saturation limit, $H$ is second order in $d/d\rho $.  The Hamiltonian $H$ has no potential, only a non-linear kinetic energy term.  It therefore describes
non-linear quantum diffusion.  Recall that the ordinary linear  diffusion equation is
\begin{equation}
  {d \over {dt}} \psi = - {p^2 \over 2} \psi
\end{equation}
It has a solution
\begin{equation}
   \psi \sim e^{-x^2/2t}
\end{equation}
The wavefunction spreads as time goes to infinity, and the exponential behaviour is universal.  It is therefore natural to expect that the solution to the JIMWLK equations will allow the saturation momentum to grow forever as $y \rightarrow \infty$, and that its solution will be universal.  This means that the Color Glass Condensate is universal and is such is fundamental.

\section{The Nature of the Color Glass Condensate Fields}

At high hadron $p^+$ , $x^-$ is a small co-ordinate and $x^+$ is big.  The components of the gluonic classical fields have the properties that  $F^{i+} $ is big, $F^{i-}$ is small, and $F^{ij}$ is of order one.  Concentrating on the large components of fields, we conclude therefore that $E \perp B \perp \hat{z}$.  The fields are therefore simply Lorentz boosted Coulomb fields.  Their distribution in color, polarization and on the two dimensional sheet corresponding to the hadron are what is determined by the theory of the Color Glass Condensate.  Note
that the density of gluons per unit area up to the saturation momentum is given by dimensional grounds as
\begin{equation}
{1 \over {\pi R^2}} {{dN} \over {dy}} \sim {1 \over \alpha_S} Q_{sat}^2
\end{equation}

It is useful to determine the light cone gauge vector potential.  We require a vector potential that gives
$F^{ij}$ and $F^{i-}$ zero everywhere and $F^{i+}$ a delta function of $x^-$ and an arbitrary function of $x_T$, independent of $x^+$.  A gauge field that is a two dimensional (in transverse coordinates) gauge transform of vacuum will give zero $F^{\mu \nu}$  If we make the vector potential different gauge transforms
of vacuum with a discontinuity at $x^- = 0$, this will give the desired form for $F^{\mu \nu}$.  We have therefore
\begin{equation}
   A^j = \theta(-x^-) {1 \over i} U_1 \nabla_T^j U^\dagger_1 + \theta(x^-)  {1 \over i} U_1 \nabla_T^j U^\dagger_1
\end{equation}
(In practice, it is sometimes necessary to spread the source out a bit in $x^-$, in a manner prescribed by the renormalization group equations.)
Notice that although the field strength $F^{\mu \nu}$ is confined to the sheet at $x^- = 0$, the Wigner distribution function corresponding to the gluon distribution function
\begin{equation}
   W^{ij}(X,p) = \int~dx^-d^2x_T ~e^{i p \cdot x} < A^i(X-x/2) A^j(X+x/2) >
\end{equation}
is spread out with a distance scale of order $1/p^+$ for components with momentum $p^+$.  The Wigner distribution is however not positive definite, and should not be expected to be so because we are measuring spatial distributions on the size scale of quantum fluctuations.

\section{The Glasma}

The collision of two sheets of Colored Glass produce color electric and magnetic fields with very different properties than those in the initial sheets.  These fields are produced in the time it takes the sheets to pass through one another, which is a very short time $t \sim e^{-\kappa/\alpha_S}/Q_{sat}$,  compared to the natural time scale for the classical fields produced after the collision $t \sim 1/Q_{sat}$.
\begin{figure}[!htb]
\begin{center}
  \mbox{{\epsfig{figure=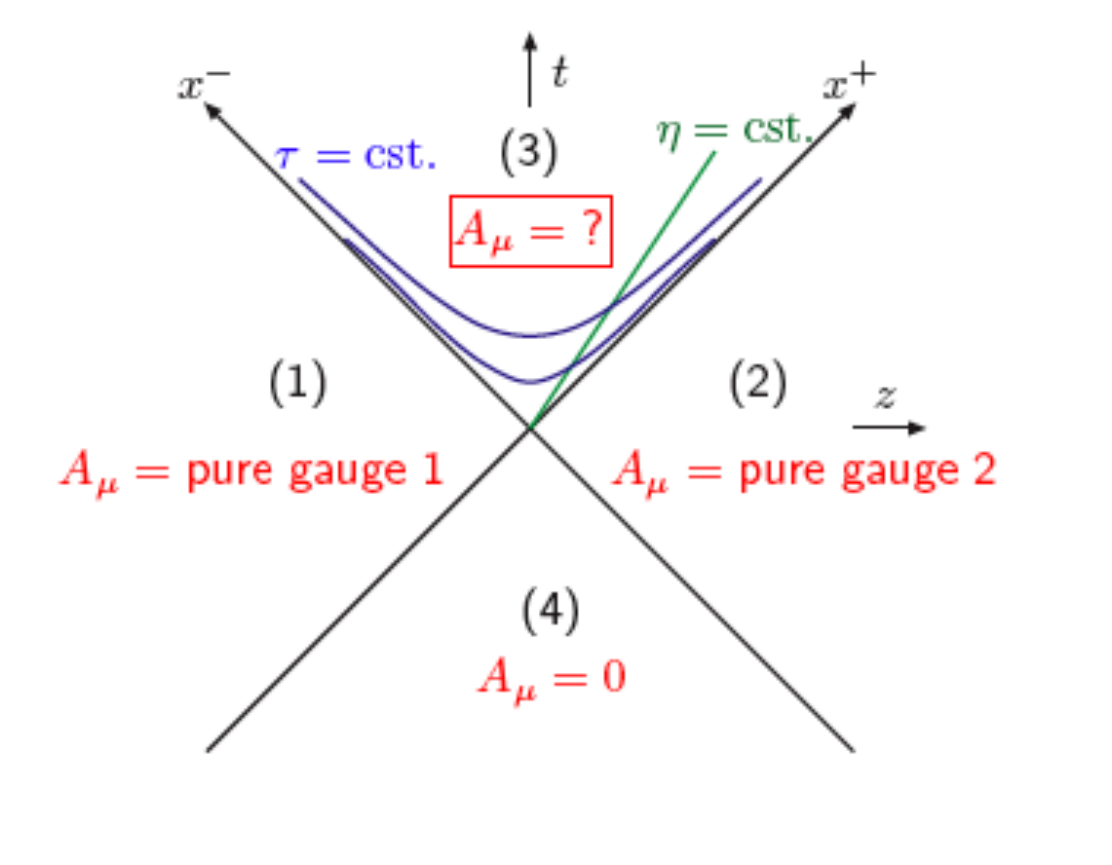, width=0.70\textwidth}}}
   \end{center}
\caption[*]{ \small  The space time diagram showing the fields before and after the collision of two sheets of Colored Glass.}
     \label{collision}
\end{figure}

In Fig. \ref{collision}, the fields are shown on a light cone diagram.  In the backward light cone there is a gauge we can choose where the fields are zero.  On the side lightcones, we choose the fields $A_1$  and $A_2$ two be two different pure gauge transforms of vacuum.  These choices give the proper field charge density on the sheets at $x^\pm = 0$ for $t < 0$.  In the forward light cone, we could try to make the charge density on the sheets be the correct value by choosing the field in the forward light cone to be $A_1 + A_2$.  This is in fact correct very close to the forward lightcone as can be seen by inspecting the Yang-Mills field equations.  The field $A_1+A_2$ is however not a gauge transform of vacuum, so the fields will evolve in the forward light cone.  One has produced a distribution of colored fields.

\begin{figure}[!htb]
\begin{center}
  \mbox{{\epsfig{figure=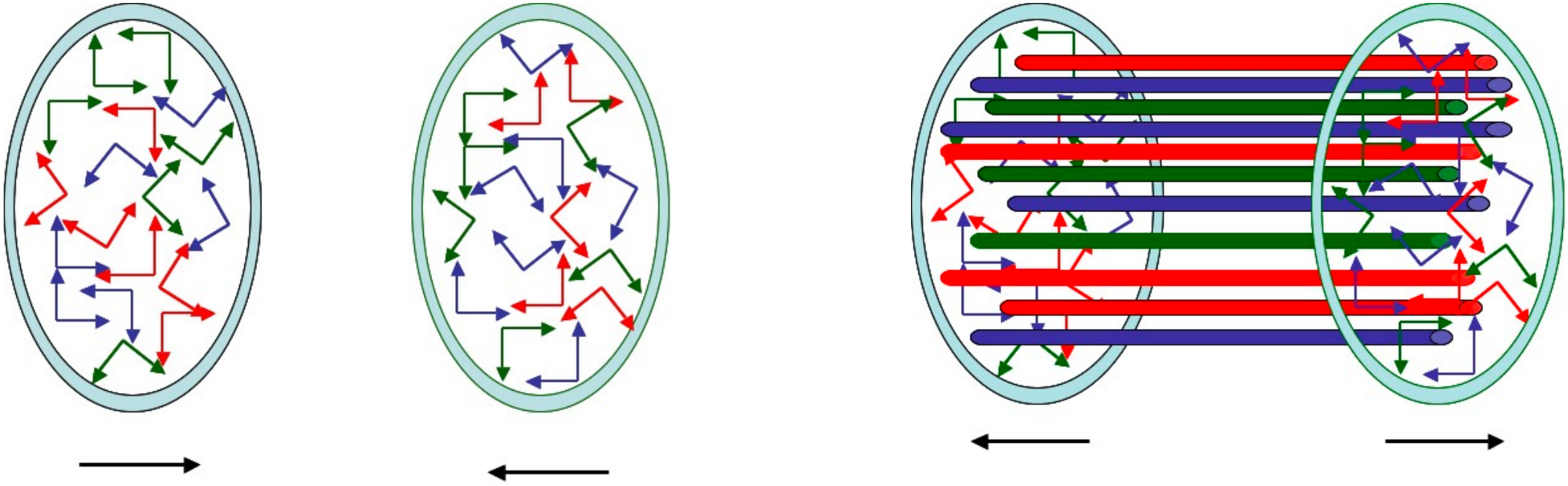, width=0.95\textwidth}}}
   \end{center}
\caption[*]{ \small  The formation of longitudinal color electric and color magnetic flux lines during the collision of two sheets of Colored Glass}
     \label{collision2}
\end{figure}

The terms in the Yang Mills equations $\vec{A} \cdot \vec{E}$ and $\vec{A} \cdot \vec{B}$ produce sources of colored electric and magnetic charge.  There are delta function contributions arising from terms like $A_i \cdot  E_j$ and $A_i \cdot B_j$ for $i \neq j$.  This means that in the collision the sheets are dusted with
equal and opposite charge densities of colored electric and colored magnetic charges.  Immediately after the collision, lines of longitudinal color electric and magnetic flux are produced.  This is shown in Fig. \ref{collision2}

The transverse density of these flux tubes is $1/Q_{sat}^2$, and the typical strength of the fields is 
$A \sim Q_{sat}/g_S$, so they are highly coherent.  there is both color electric and color magnetic field,
so there is a large topological charge density $\vec{E} \cdot \vec{B}$.  The fields evolve and decay according to the classical Yang-Mills equations.  Unlike the case where there is pair production needed for the decay of an electric flux tube, because there is both a color electric and a color magnetic field, the Yang-Mills equation allow for the classical decay.

\section{Phenomenology of the Color Glass Condensate and the Glasma}

There are now a wide variety of phenomenon described by the Color Glass Condensate and the Glasma.  The Color Glass Condensate provides a good description of deep inelastic scattering and diffraction in electron-proton collisions at HERA for small values of $x$.  It has provided a good phenomenological description of heavy ion collisions and $d Au$ collisions at the RHIC accelerator.  An excellent rveiew of the situation is provided at www.bnl.gov/riken/glasma (to be published in a special edition of Nuclear Physics A).
An even better summary of the current situation is provided in the presentation of J. P. Blaizot.  Of course at the LHC, the ideas of gluon saturation can and will be tested in pp, pA and AA collisions.  At the LHC, $x$ values are very small and the saturation momentum large enough so  that computations should become precise.  There is also the possibility to test these ideas in the collisions of electrons from nuclei at an electron-heavy ion collider such as eRHIC.

\section*{Acknowledgments}

 It is an honor to speak at a memorial meeting for a truly gifted, creative and absolutely honest physicist: Volodya Gribov  His work has influenced me in many ways including some that I have only come to understand with time.
 
I thank Luciano Bertocchi, Yuri Dokshitzer, Peter Levai, Julia Nyiri,
and Daniele Treleani for organizing this meeting in honor of Volodya Gribov and the sponsors, ICTP (Trieste), RMKI (Budapest) and INFN. 
The research of L. McLerran is supported under DOE Contract No. DE-AC02-98CH10886.

\end{document}